\documentclass[12pt, journal, draftcls, onecolumn]{IEEEtran}
\usepackage{algorithm}
\usepackage{algorithmic}
\usepackage{amsfonts}
\usepackage{amssymb}
\usepackage{cite}
\usepackage{color}
\usepackage{multirow}
\usepackage{subfigure}
\usepackage{graphicx,times,amsmath}

\begin{document}

\title{Configurations and Diagnosis for Ultra-Dense Heterogeneous Networks: From Empirical Measurements to Technical Solutions}

\author{Wei~Wang,~\IEEEmembership{Member,~IEEE}, Lin~Yang, Qian~Zhang,~\IEEEmembership{Fellow,~IEEE}, Tao~Jiang,~\IEEEmembership{Senior Member,~IEEE}
\thanks{W. Wang and T. Jiang are with the School of Electronic Information and Communications, Huazhong University of Science and Technology. L. Yang and Q. Zhang are with the Department of Computer Science and Engineering, Hong Kong University of Science and Technology, Hong Kong. Corresponding author: Tao Jiang.}}

\maketitle

\begin{abstract}
The intense demands for higher data rates and ubiquitous network coverage have raised the stakes on developing new network topology and architecture to meet these ever-increasing demands in a cost-effective manner. The telecommunication industry and international standardization bodies have placed considerable attention to the deployment of ultra-dense heterogeneous small-scale cells over existing cellular systems. Those small-scale cells, although provide higher data rates and better indoor coverage by reducing the distance between base stations (BSs) and end users, have raised severe configuration concerns. As the deployments are becoming irregular and flexible, inappropriate configurations occur frequently and undermine the network reliability and service quality. We envision that the fine-grained characterization of user traffic is a key pillar to diagnosing inappropriate configurations. In this article, we investigate the fine-grained traffic patterns of mobile users by analyzing the network data containing millions of subscribers and covering thousands of cells in a large metropolitan area. We characterize traffic patterns and mobility behaviors of users and geospatial properties of cells, and discuss how the heterogeneity of these characteristics affects network configurations and diagnosis in future ultra-dense small cells. Based on these observations from our measurements, we investigate possible models and corresponding challenges, and propose a heterogeneity-aware scheme that takes into account the disparity of user mobility behaviors and geospatial properties among small cells.
\end{abstract}

\section{Introduction}
The ever growing data demands and ubiquitous access from users of new generation mobile Internet devices, such as smartphones and tablets, are driving current mobile communication systems to their limits. Wireless operators are urgently seeking cost-effective solutions to boost network capacity and coverage. A promising approach to this problem is the deployment of smaller cell structures, including picocells, femtocells and so on. These small cells operate using the same spectrum and technology as traditional macrocells in existing cellular networks. Compared to traditional macrocells, small cells largely shorten the distance between base stations (BSs) and user terminals (UEs), thereby offering better indoor coverage and higher capacity with increased spatial reuse~\cite{ghosh2012heterogeneous}. Millions of commercial femtocells and picocells have already been deployed by major operators worldwide, including Sprint, Verizon, AT$\&$T in USA, Vodafone in Europe, and Softbank in Japan. 

On the one hand, the deployment of heterogeneous small cells reaps the benefits of spatial reuse to increase the capacity of cellular networks in dense areas with high traffic demands~\cite{peng2014heterogeneous}. On the other hand, these heterogeneous small cells make network topology increasingly complicated and irregular, which has imposed wide network configuration concerns. Such configuration issues come from the fact that inappropriate configurations (such as transmission power, channels, and handover parameters) of one small cell can lead to severe performance degradation or even outage of a set of surrounding cells, as these cells are densely deployed with a large portion of overlapped coverage. While most existing work has gone into the initial configuration and network optimization, there have been far fewer efforts made on configuration diagnosis. Configuration diagnosis aims at determining the causes of failure or poor performance due to inappropriate configurations by analyzing network measurements. The need for configuration diagnosis comes from the fact that since network topologies have become more complicated and irregular, it is hard to precisely model the relation between the network performance and configurations without certain assumptions.

Typically, ultra-dense small cells are designed to cover dozens of meters or less and each cell serves only a couple of users. Human activities, usage behavior and perceived experience of users weigh increasingly on the performance of a cell due to cell densification. Fine-grained user behaviors are envisioned to facilitate configuration and diagnosis in ultra-dense small-cell networks. In particular, whether a configuration is appropriate depends on user's application-specific traffic patterns, mobility behaviors and their relations with geospatial properties. In addition, human activities and usage behavior are heterogeneous across different cells.  For example, video traffic volume in residential areas is normally larger than that in office areas. Therefore, user heterogeneity should be taken into account when configuring and diagnosing different small cells.

The goal of this article is to first investigate the fine-grained user behaviors in cellular networks through empirical measurements, and call attention to a clean-slate redesign of configuration diagnosis frameworks for ultra-dense small cell networks. We start by taking a deep dive at analyzing application usage behaviors of cellular users in one large metropolitan area, and then discuss their impact on future ultra-dense heterogeneous small cell networks (ultra-dense HetNets). Different configuration diagnosis models and their challenges are discussed. We shed light on a heterogeneity-aware framework that exploits heterogeneous usage behaviors across cells. 

\section{Exploiting User Heterogeneity in Ultra-Dense HetNets: Traffic, Mobility, and Distribution}
\subsection{Development and System Architecture of Ultra-Dense HetNets}

The traditional cellular systems (1G-3G systems) consist of large-scale, standalone macro-cells where each cell operates independently and intercell interference is avoided by employing static frequency planning or code-division multiple access (CDMA) techniques~\cite{peng2014heterogeneous}. 4G systems improves spectrum efficiency by reusing frequency in adjacent cells, at the cost of introducing intercell interference. These cellular systems are designed from the coverage perspective by deploying cells at different sites to serve both indoor and outdoor areas with minimum quality of service. With the rapid growth in the number of devices accessing networks, the bottleneck of cellular systems transits from the coverage limited to the capacity limited state. To further improve spectrum efficiency and network capacity, existing cellular systems are envisioned to be embedded with multiple tiers of smaller cells including micro-cells, pico-cells, femto-cells, thereby forming a hierarchical cellular structure where large-scale cells guarantee coverage while small-scale cells are deployed in local hotspot areas to match the capacity demands~\cite{liu2015device}.

\subsection{Understanding User Heterogeneity}
Conventionally, independent homogeneous Poisson point processes are used to model user and BS distributions~\cite{tabassum2014interference,bai2013location}. While this model is suitable for modeling traditional cellular systems, while it falls inaccurate in ultra-dense HetNets. As small cells are deployed with the consideration of local capacity demands~\cite{andrews2014overview}, the deployments of small cells and user or traffic distribution are correlated. In addition, the average intercell distance is merely dozens of meters and each cell supports only a few users. Compared to traditional large-scale cells, the impact of heterogeneous user and traffic distributions is amplified in small cells. The user heterogeneity has been taken into account at a high level in modeling HetNets~\cite{mirahsan2015user,schoenen2014user}. We take one step further to consider fine-grained user heterogeneity at the application-level. To this end, we conduct an empirical study on large-scale anonymized IP flow traces from a tier-1 cellular provider in China, and envision the impact of heterogeneity in future ultra-dense HetNets. {It is worth noting that the adoption of small cells is envisioned in future 5G, which has not yet been deployed in today's cellular systems. To extract useful information from the measurements in today's cellular networks, we exploit human behaviors in terms of mobility, spatial, and app usage patterns, which largely depend on points of interests and population distributions in cities. These observations are independent of network architectures, and hence are applicable to provide some hints for future architecture design.}

\subsubsection{Experiment Configurations}
{We collect cellular traces from a tier-1 cellular service provider in China. {Details of the test parameters and data set are summarized in Table~\ref{tab:dataset}. {We study the behaviors of more than 8 million users in a large metropolitan area, where there are 13,120 BSs in a 24~km $\times$ 22~km metropolitan area, which includes different types of regions, such as train stations, an airport, educational institutions, business districts, government office area, shopping malls, and so on. The average received signal strength indicator (RSSI) of all BSs is -84.8~dBm, while the average RSSI of serving BS is -~71.2~dBm. Detailed RSSI statistics are listed in Table~\ref{tab:dataset}. The average distance between BSs is 277.5~m. The average speed of users is 1.5~m/s, which is computed from the location of cells and user traffic. There are mainly three types of user movements: in the stationary state, walking/running, and in vehicle.}} The data is imported from the operation supporting systems (OSS) in cellular networks. Our data collection platform acquires the OSS data by probing and interpreting the data to x-Detail Record (xDR) tables including User Fkiw Detail Record (UFDR), Transaction Detail Record (TDR), and Statistics Detail Record (SDR), which include International Mobile Subscriber Identification (IMSI) or International Mobile Equipment Identity (IMEI). IMSI and IMEI are used to identify data corresponding to each user. xDR tables are stored in Hadoop distributed file systems (HDFS) that communicates with Hive/Spark SQL for data processing.}

{{We continuously collect the OSS data in 3G/4G networks of one service provider for 13 consecutive days.} The data record of each day is about 800~GB. We analyze all the IP-layer flows carried in the Packet Data Protocol (PDP) context tunnels, \textit{i.e.}, flows that are sent to and from mobile devices. Each IP flow contains application information and location information. The application information consists of application ID, application name (e.g., Google Maps), protocol, source IP address \& port, destination IP address \& port, time spent on transmission and transmission speed. The location information contains the Mobile Country Code (MCC), Mobile Network Code (MNC), Location Area Code (LAC) and Sector ID (SI). This location information is obtained by joining the PDP sessions with a fine-grained log of signaling messages, which includes detailed logs of handover events.}

\begin{table}[!htb]
	\begin{center}
		\caption{Summary of test parameters and data set}
		\label{tab:dataset}
		\begin{tabular}{|c|c||c|c|}
			\hline Average distance to srv BS & 187~m & Median distance to srv BS & 157~m\\
			\hline
			Average RSSI & -84.8~dBm & Median RSSI & -74.6~dBm\\
			\hline
			Average RSSI of srv BS & -71.2~dBm & Median RSSI of srv BS & -64.3~dBm\\
			\hline
			Average distance between BSs & 277.5~m & Area & 24~km $\times$ 22~km\\
			\hline
			\hline Duration & 13 days & Data volume & 800~GB/day\\
			\hline Sectors & 38360 & Active subscribers & 8,000,000 \\
			\hline\hline
			\textbf{Category} & \textbf{\# apps} & \textbf{Category} & \textbf{\# apps}\\
			\hline
			web browsing & 6 & P2P & 9 \\
			\hline
			instant message & 6 & reading & 16 \\
			\hline
			social networks & 5 & video & 9 \\
			\hline
			music & 16 & app market & 3\\
			\hline
			game & 70 & email & 4\\
			\hline
			stock trading & 8 & online shopping & 6\\
			\hline
			map & 2 &  &  \\
			\hline
		\end{tabular}
	\end{center}
	\vspace{-0.5cm}
\end{table} 

{We preprocess the raw IP flows to categorize traffic into different types. We adopt the Deep Packet Inspection (DPI) solution provided by the cellular service provider to identify the application information by examining the various fields such as IP and port in the packet header or even the data content. Additionally, some heuristic rules and advanced pattern mining techniques are employed to learn application information from the consecutive traffic pattern during a session.} Furthermore, we manually categorize these applications into 13 groups according to their function or genre: (1) \emph{web browsing (WEB)} (2) \emph{P2P}; (3) \emph{instant message (IM)}; (4) \emph{reading (RE)} includes news reading apps; (5) \emph{social network (SN)}; (6) \emph{video (VD)}; (7) \emph{music (MU)}; (8) \emph{app market (AM)}; (9) \emph{game (GM)}; (10) \emph{email (EM)}; (11) \emph{stock trading (ST)}; (12) \emph{online shopping (SH)}; (13) \emph{maps}. After that, accurate location information, \textit{e.g.}, Location Area Code (LAC) and Sector ID (SI), is obtained by joining PDP sessions with a fine-grained log of signaling messages, which has detailed records of all handover events.

\begin{figure*}
	\centering
	\includegraphics[width=0.8\textwidth]{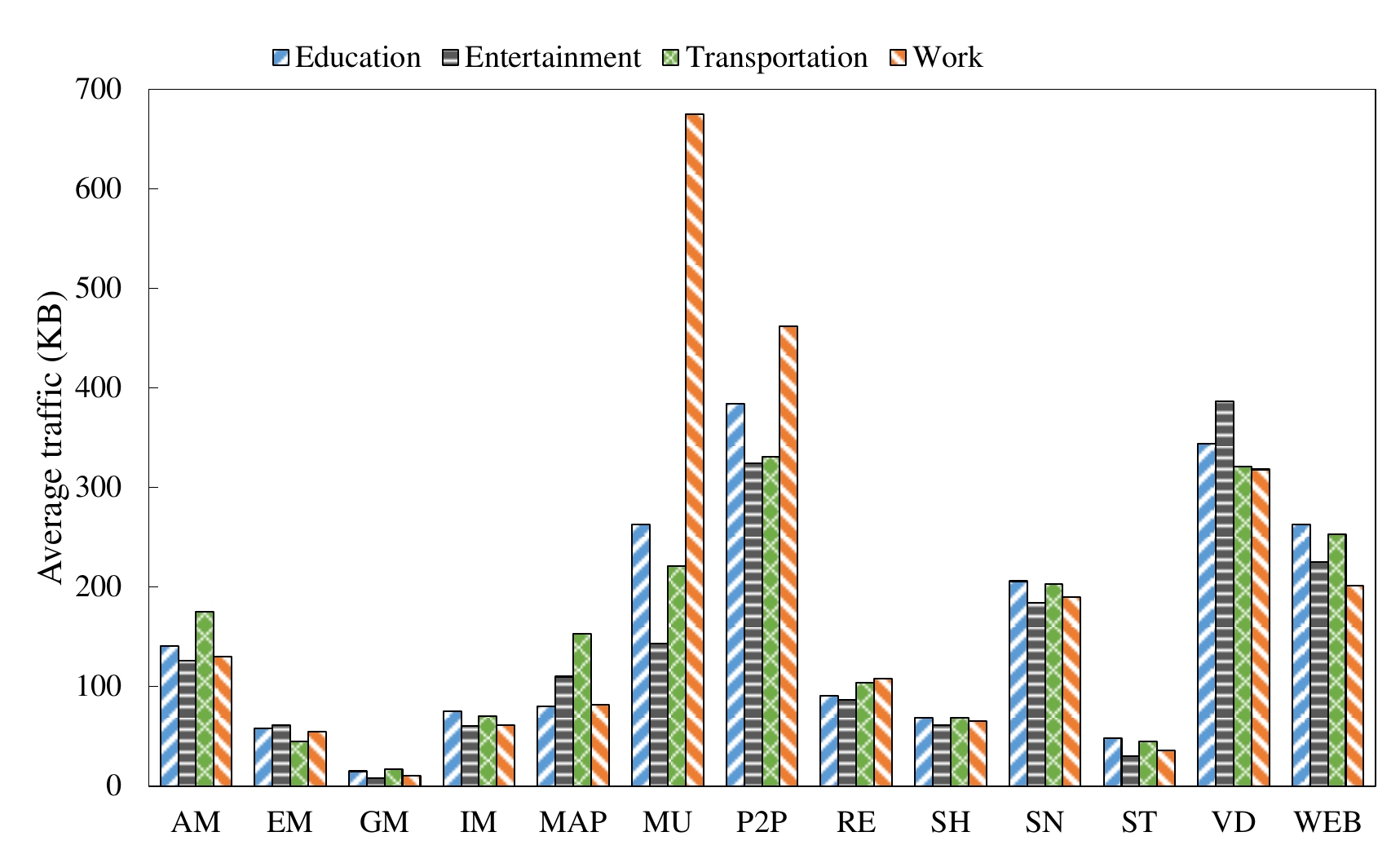}
	\caption{Application usage in cells of different covering areas. This indicates the location type can significantly affect users' application usage pattern.}
	\label{fig_spatial_pattern}
\end{figure*}

\begin{figure*}
	{\label{fig_ht_traffic_contrib_over_location_normal} 
	\includegraphics[width=3.5in]{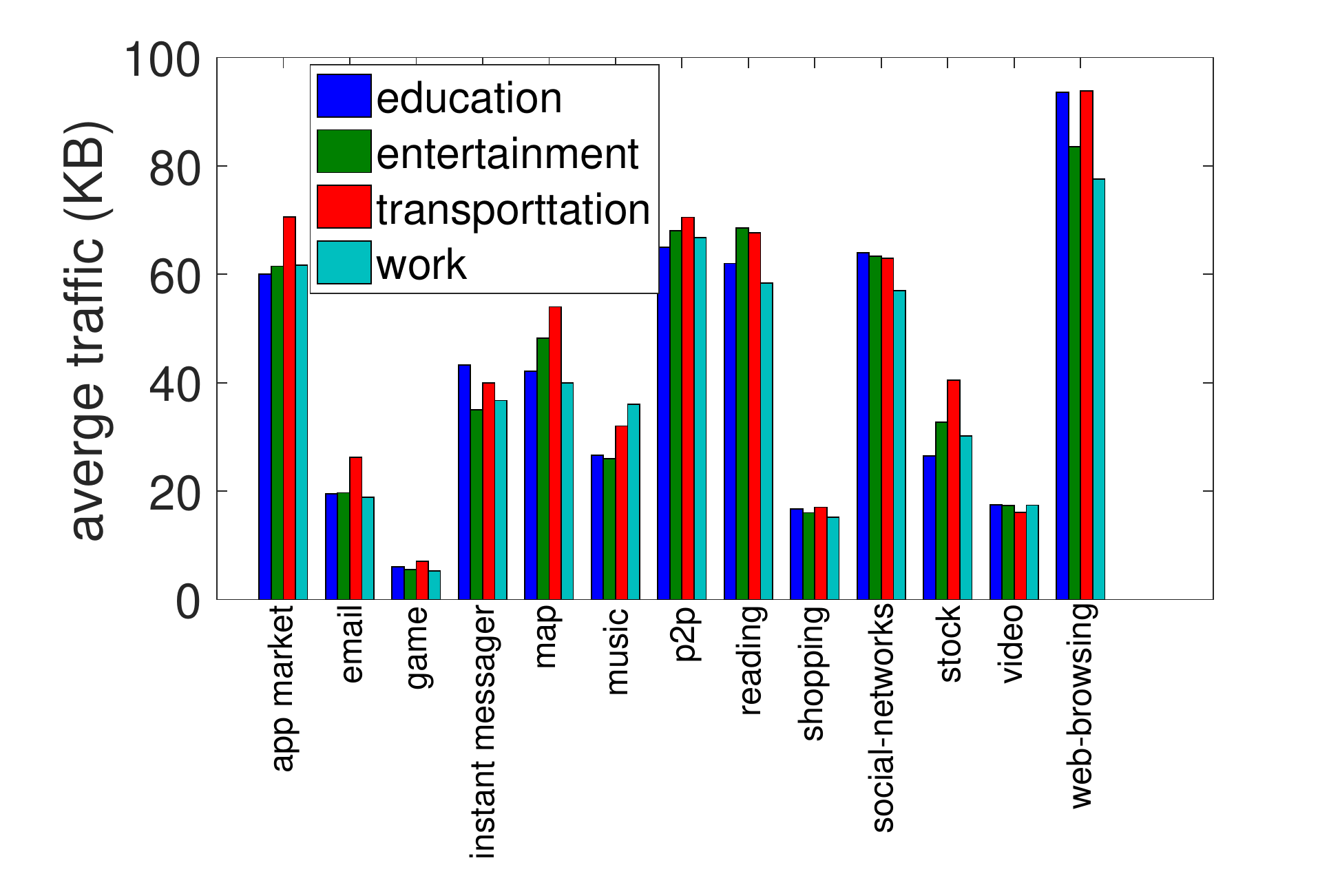}}
\hspace{-0.67cm}
{\label{fig_ht_traffic_contrib_over_location_heavy} 
	\includegraphics[width=3.5in]{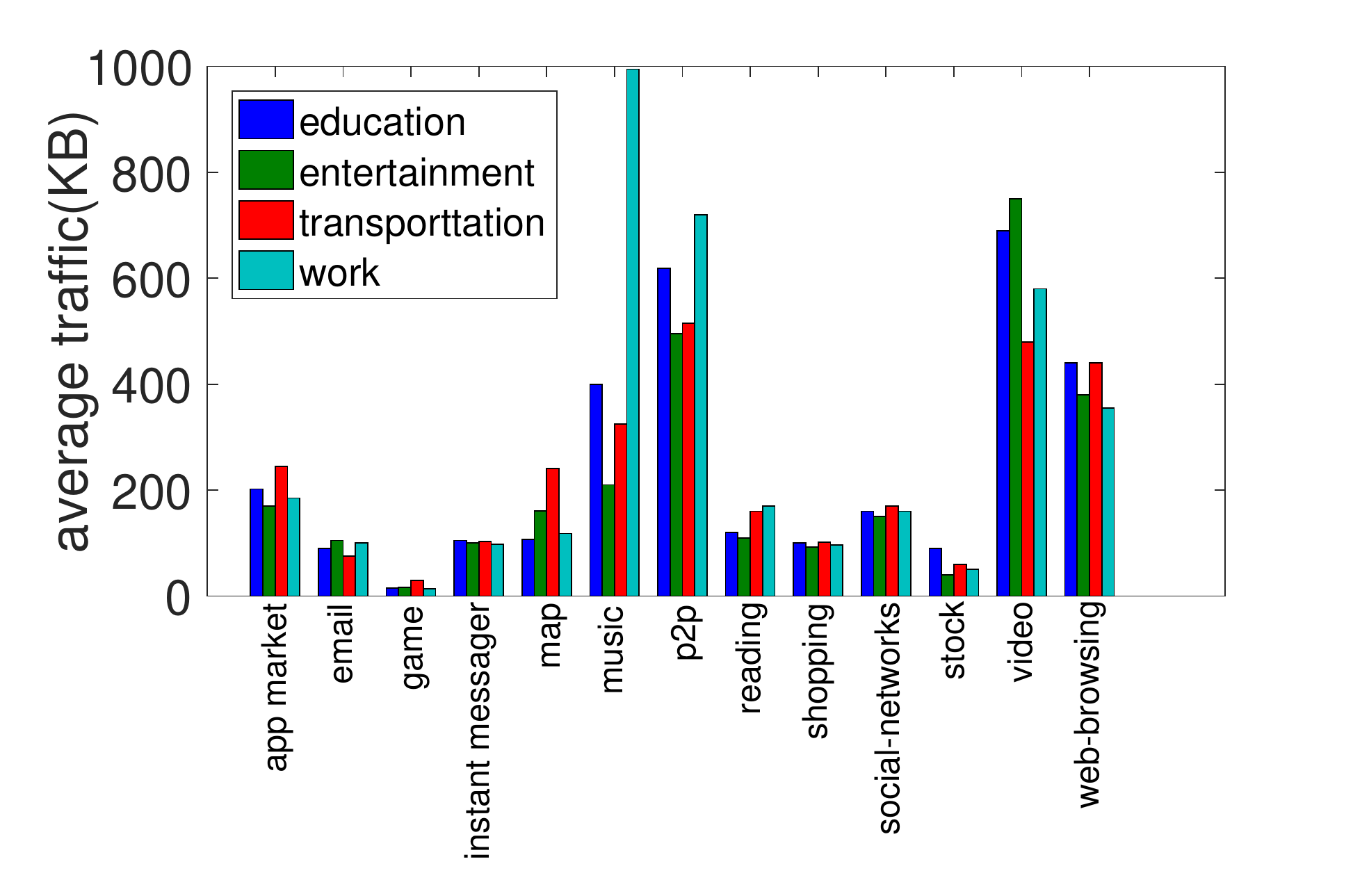}}
\caption{Impact of location scenarios on network usage behaviors. The left (right) figure shows the average traffic for normal (heavy) traffic users.}
\label{fig_ht_app_usage_over_location}
\end{figure*}

\subsubsection{Spatial Pattern}
We begin our analysis with the application usage in cells of different covering areas. As various applications pose distinctive requirements on the underlying network service, difference in application usage patterns can imply the discrepancies in network configuration and quality requirement.

To take a fine-grained look at the spatial patterns of application usage, we first estimate the coverage of each cell according to its distance to nearby cells and heuristically classify each cell into four types according to the function of its covering area: \textbf{(a) Transportation}, such as train stations or airports; \textbf{(b) Educational institution} includes schools, colleges and research institutions; \textbf{(c) Work} consists of business districts, campuses of large corporations and government office areas; \textbf{(d) Entertainment} covers large shopping malls and places of interest in the downtown. In total, we have identified the covering type of 1241 cells in our data set and all these location types are manually validated one by one. 

In Figure~\ref{fig_spatial_pattern}, we compare the average traffic generated from each application category in cells of different types. We can observe that the traffic patterns of some applications vary with locations. For example, \textit{music} dominates in \textit{work} areas, \textit{P2P} is more popular in \textit{education} and \textit{work} areas than others, while \textit{map} contributes the highest traffic in \textit{transportation} areas. These trends indicate that location can affect the extent to which an application is used. Also, as various applications pose different traffic patterns and network quality requirements, we can expect distinctive traffic patterns for the cells covering different areas. For instance, a cell mainly covering \textit{work} areas may pose a high requirement on bandwidth as \textit{music} and \textit{P2P} applications are frequently used in this area. 

{We further examine the impact of different location scenarios on users' network usage behaviors. Figure~\ref{fig_ht_app_usage_over_location} shows that scenarios significantly affect the app traffic of heavy traffic users while imposing little impact on normal traffic users. The results imply that user heterogeneity as well as location scenarios should be considered when diagnosing network configurations.}

\subsubsection{Mobility Pattern}
Another significant user heterogeneity is their mobility pattern. We wonder is there any difference in users' network requirements as their mobility patterns are distinct. {To quantify user's mobility, we use two metrics: i) \textit{number of visited cells}, which is the number of cells visited by a user in one day, and ii) \textit{radius of gyration (RoG)}, which is commonly used in the study of human mobility~\cite{gonzalez2008understanding} and can be interpreted as the geographical area traveled by a user.}

\begin{small}
	\begin{equation}
	\label{eq_rog}
	r_g = \sqrt{ \frac{1}{n} \sum\limits_{i=1}^{n}{(l_i - l_{mass})^2} },
	\end{equation}
\end{small}
{where $ l_i$ is the latitude and longitude of cell $ i $, $ l_{mass} = \frac{1}{n}\sum\limits_{i=1}^{n}{l_i} $ is the center of mass of a user's trajectory and $ (l_i-l_{mass}) $ is the Euclidean distance between $ l_i $ and $ l_{mass} $. The number of cells measures a user's trajectory length in terms of cells, while RoG represents the geospatial coverage of a user's mobility.}


\begin{figure*}
	\centering
	{
		\label{fig_ht_per_capita_traffic_over_mobility_normal} 
		\includegraphics[width=0.95\textwidth]{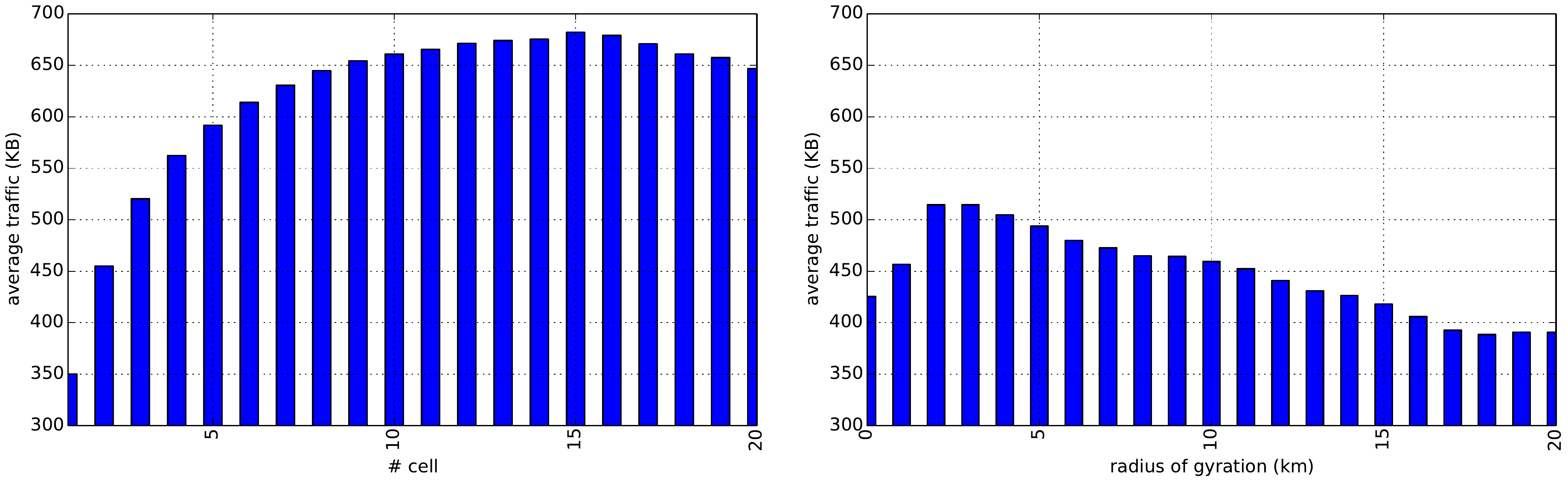}}
	\hspace{0.03in}
	{
		\label{fig_ht_per_capita_traffic_over_mobility_heavy} 
		\includegraphics[width=0.95\textwidth]{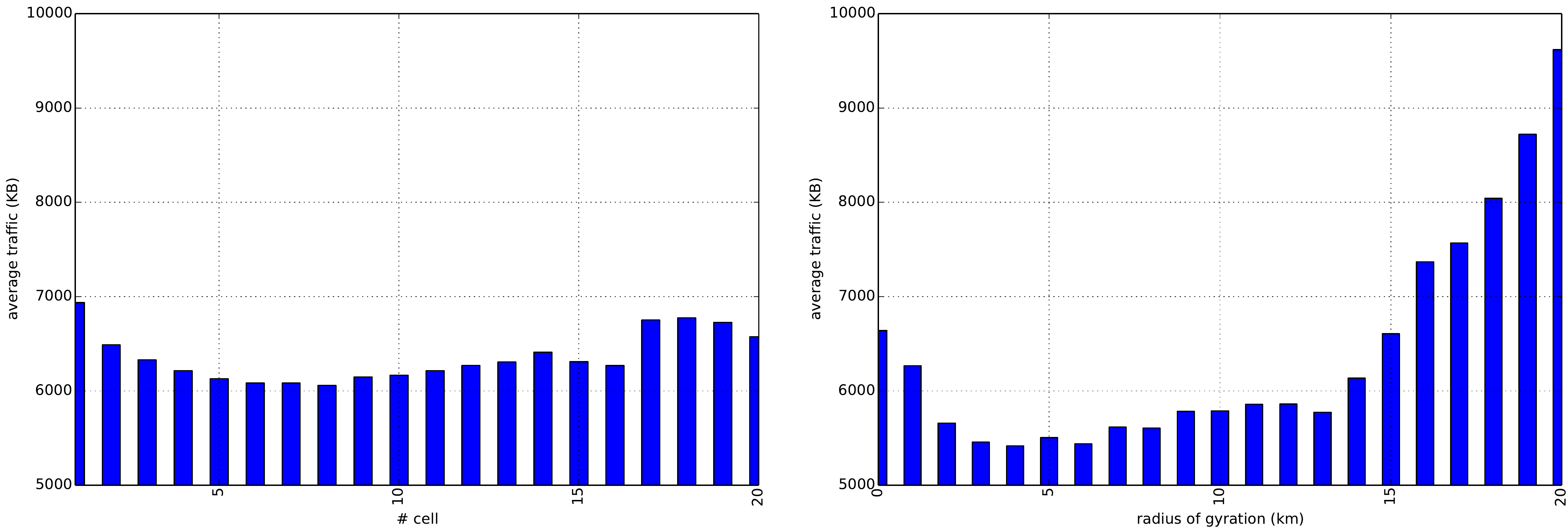}}
	\caption{Impact of mobility on traffic. Top and bottom figures are the results for normal and heavy traffic users, respectively.}
	\label{fig_mobility}
\end{figure*}


{Figure~\ref{fig_mobility} shows the impact of mobility on users of different traffic volumes. Interestingly, we observe that the average traffic of normal traffic users increases dramatically with the number of visited cells, while showing completely different trends in terms of RoG. This implies that the normal traffic users tend to generate more traffic when they are moving within a small region.}

\begin{figure}[t]
	\centering
	\includegraphics[width=6.5in]{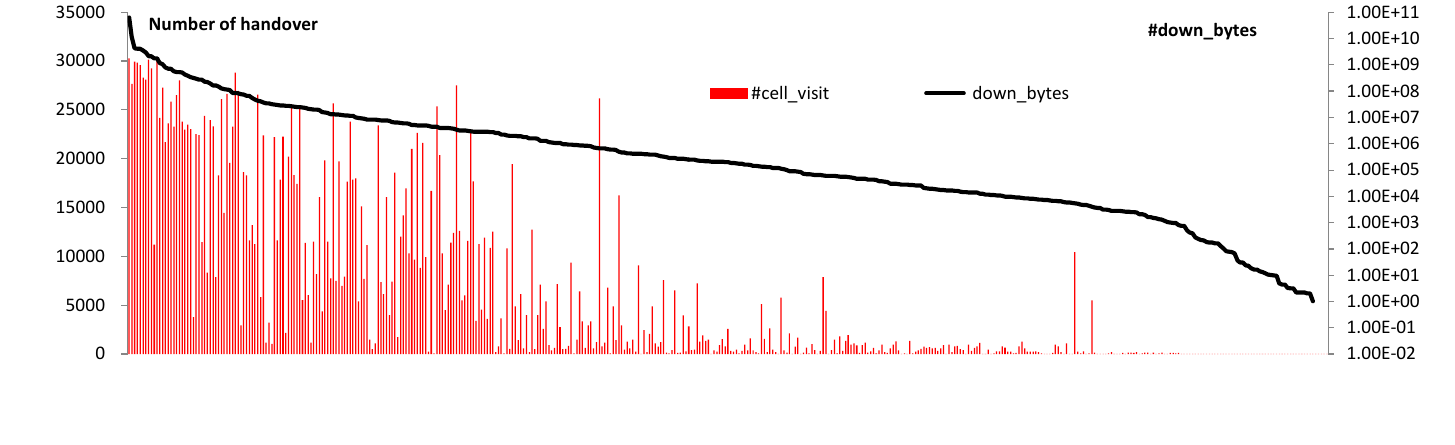} 
	\caption{Impact of handover on traffic volume.}
	\label{fig:handover}
\end{figure}

{We also measure the number of handover events from the fine-grained signaling messages in the OSS data. We observe that although a single handover event alone may not have an immediate impact on user behaviors, the frequency of handover events, that is, the number of handover events per day, highly correlate with traffic volume. Figure~\ref{fig:handover} shows the number of handover events and traffic volume that correspond to each user in the OSS data. The results imply that frequent handover events correlate with large traffic volume. The reason is that more frequency handover events occur in scenarios of trains, buses, taxis, where users use their smartphones to surf the Internet, while minimal handover events occur in scenarios of offices and homes where users consume most of their data using Wi-Fi instead of cellular. Therefore, handover patterns are highly related to scenarios and user's network behaviors.}

These trends indicates that the diversity in user mobility can introduce significant user heterogeneity by affecting their traffic patterns and application usage.

\subsubsection{The Vision of Heterogeneity in Future Ultra-Dense HetNets}
It is envisioned that today's cellular systems would continue evolving to an extremely dense network of heterogeneous small cells, whose number approaches that of mobile users. Such network densification will enlarge the heterogeneity among cells, in that (i) small cells are deployed in different hotspot areas where usage behaviors can be quite different, and (ii) the performance of small cells is more sensitive to the usage behaviors of individual users due to the scarcity of users in one cell. Consequently, the impact of heterogeneous traffic patterns, mobility behaviors and geospatial properties is amplified in small cells, and thus the awareness of the heterogeneity among cells is a key pillar when configuring and diagnosing ultra-dense HetNets.

\section{Configuring and Diagnosing Ultra-Dense HetNets: Models and Challenges}

Configuration and diagnosis aim at setting up cells with proper configurations and identifying bad parameter settings (cause diagnosis). Normally, multiple performance indicators and counters are used as criteria. These indicators and counters are statistical features of field measurements such as blocked/dropped call rate (BCR/DCR), Block Error Rate (BLER), number of handover attempts and so on. {Normally, a handover event is triggered when (i) the difference in received signal power values from the serving cell and a neighboring cell is smaller than a threshold, known as the \textit{handover margin}; and (ii) the signal to interference and noise ratio (SINR) is persistently lower than a threshold for a period of time referred to as Time-to-Trigger (TTT).} {In HetNets, handover events are divided into two types, that are, horizontal handover and vertical handover. Horizontal handover refers to the handover between cells in the same tier while vertical handover refers to the handover across different tiers, such as handover between a femtocell and a macrocell.} {High mobility users with more frequent handover events tend to have higher BCR/DCR, larger variance of RSSI, and lower voice quality, as it easily leads to early or late handover that will result in QoS drop. Hence, the handover parameters need to be carefully tuned for different scenarios.}

Operators have proposed to standardize measurement collection and monitoring in 3GPP. The Release 9 version of the 3GPP specifications~\cite{EUTRAN2} introduced mobility robustness optimisation (MRO) to help detect and correct mobility related misconfigurations, and categorized handover related failures into three types: too late handover triggering, too early handover triggering, and handover to a wrong cell. {These indicators are used to assess the performance of current handover margin and TTT settings.} A work item named Minimization of Drive Tests (MDT) was created in 3GPP's Release 10 specification to collect end user's radio measurements during user's idle state~\cite{MDTmag}. The abnormality of these performance indicators is referred to as \textit{symptom} in configuration and diagnosis. These performance indicators are logged and reported to Operations, Administration and Management (OAM) servers for periodic drive tests. Taking these indicators as input, a configuration and diagnosis mechanism identify proper and improper configurations via a diagnostic model, which is normally built based on historical \textit{instances}, which consists of measurements or indicators together with the corresponding configurations.

To extract the latent relations between the performance indicators and configurations, the state-of-the-art systems turn to statistical modeling or machine learning techniques. These systems are mainly based on inference or classification models that are constructed based on historical instances collected in the network. However, the historical instances for small cells are scarce due to the limited number of users in each cell. 
\subsection{Standalone Model}

As the number of small cells are massive and some are even installed on the consumer's premises, it is natural and efficient to manage them in a distributed and self-organizing manner~\cite{peng2013self,wang2014cod}, in which each BS sets up and tunes its configuration based on the measurements reported from its associated users. To facilitate configuration and diagnosis that align to the distributed manner, an intuitive method is to allow each small cell to run the configuration mechanism independently based on the cell-specific measurements. In such an architecture, the configuration model for each cell is constructed independently based on the field measurements collected by the BS.

Ideally, this standalone model tailors configurations to each cell according to the local wireless environments and traffic patterns. It requires the collection of a large amount of measurement data to train accurate detection and diagnostic models, which, however, can hardly be satisfied by the cell-specific architecture. Since the cell sizes in ultra-dense HetNets are several orders of magnitude smaller than that of a traditional macrocell, and each small cell supports only a couple of users, the performance indicators and counters computed based on such small-scale measurements fall inaccurate to reflect the statistical features due to the sparsity of measurements with high uncertainty. Since small-scale measurements are sensitive to single user activities and variances of wireless environments like shadow fading, it requires months or even years to collect enough fault instances to construct an accurate troubleshooting model. In addition, as more and more small cells are installed, the situation where new neighboring cells join the network often occur. These deployment changes make historical measurements easily outdated and further limit the amount of measurement data for model training. 
\subsection{Unified Model}
To cope with the measurement scarcity issue, an intuitive approach is to utilize measurements across multiple cells to construct a unified configuration model. Most of today's mobile operators promote the joint deployment framework, where FBSs are installed by the users while some system parameters are controlled by the operator' OAM server~\cite{wang2014cod,lin2011macro}. This joint deployment framework makes it practically feasible to collect measurements from multiple cells within a region via backhaul connections to the operator's OAM server. In this unified model based architecture, the OAM server collects measurements from multiple cells and extracts a unified troubleshooting model for all cells.

On the one hand, the unified model seems to overcome the measurement scarcity issue faced by the standalone model. On the other hand, small cells are deployed in very different wireless environments, as observed in our empirical results. Measurements from cells with very different environments can barely help with each other, but rather be misleading in the model construction. 

\section{Heterogeneity-Aware HetNets Configurations and Diagnosis}



As both standalone and unified models have intrinsic limitations in applying to ultra-dense HetNets, we consider an advanced model that falls somewhere in between to exploit the benefits of the two yet overcome their drawbacks. To achieve this goal, we leverage the notion of transfer learning~\cite{tlsurvey} to seek proper solutions. The need for transfer learning arises when the data can be easily outdated, or from another task domain which has different features and distributions. For data from similar tasks, there must be some latent knowledge or common structure which can still be leveraged. In such cases, the data can be ``transferred'' to extract the useful knowledge. We leverage transferable knowledge learned from different cells to assist in the configuration model construction for one cell. Specifically, each cell can be considered as a task domain where the set of possible configurations and symptoms are the same for all cells. Thus, the configuration knowledge from other cells can be leveraged to build the troubleshooting model.

The transfer learning architecture aims to extract useful knowledge and structure from other cells and feed them into the model training, while filtering out misleading information. The OAM server collects the measurements from multiple cells within a region in the same way as the unified model based architecture. Then, the measurements are split into two parts: the measurements from the target cell and the rest. The measurements from the target cell are used to filter out a subset of the remaining measurements that are collected in very different environments. This is normally achieved by automatically adjusting the weights of each instance in the model training process. The measurements with large weights are considered as transferable training instances, and are used to improve the confidence of the trained model. 

{When transferring the historical data from other cells, we have the following observations. On the one hand, cells with similar traffic patterns and similar neighboring layouts share sufficient latent knowledge with the target cell. On the other hand, cells with very different wireless environments can even transfer misleading knowledge to the target cell, which is destructive rather than constructive to the diagnosis model construction.} Thus, to tailor the transfer learning algorithm to our scenario, the transfer scheme should be cell-aware, that is, the data from different cells should be considered separately to let the data from different cells have different levels of impact on the model construction. In particular, we first estimate the distribution difference of two cells' measurement data using Kullback-Leibler (K-L) divergence. As there are multiple dimensions in the measurement data, we weight each dimension according to information gain, which is defined as:
\begin{equation}\label{f:gain}
\begin{split}
\text{InfoGain}(A_i) = H(I) - H(I|A_i),
\end{split}
\end{equation}
where $H(I)$ denotes the entropy of the misconfiguration instances, and $H(I|A_i)$ denotes the condition entropy of misconfiguration instances given one dimension $A_i$ of the measurement data. If $\text{InfoGain}(A_i)$ is larger, $A_i$ is more decisive to the diagnosis results. Based on the information gain of each dimension, the divergence between two cells is computed by summing up the weighted divergences of all dimensions in the measurements. After deriving the divergences between cells, we divide the instances by cells and learn a cell-specific model for each cell. Each model is constructed based on the instances of the corresponding cell. We weight the instances of each cell according to the divergence between that cell and the target cell. As such, instances from similar cells have strong impact on the model construction. Then, we set weights for all cell-specific models. The instances from the target cell are fed into other cells' models for misconfiguration diagnosis. After weighting each model, all models are treated as voters, which are weighted by $\mathbf{\omega}$. The final result is the configuration or misconfiguration that wins the most votes.

\begin{table}[h]
	\centering
	\caption{Diagnosis accuracy under various cell density.} \label{tab:sim}
	\begin{center}
		\begin{tabular}{l|ccccc}
			\hline  Cell density  & 20 & 40 & 60 & 80 & 100 \\\hline
			Cell-specific (TX Power)  & 0.75 & 0.78 & 0.79 & 0.80 & 0.80 \\
			Unified model (TX Power) & 0.55 & 0.68 & 0.75 & 0.78 & 0.81 \\
			Transfer learning (TX Power) & 0.81 & 0.88 & 0.91 & 0.91 & 0.92 \\
			Cell-specific (HO Margin) & 0.59 & 0.58 & 0.57 & 0.60 & 0.60 \\
			Unified model (HO Margin) & 0.41 & 0.54 & 0.57 & 0.62 & 0.63 \\
			Transfer learning (HO Margin) & 0.73 & 0.82 & 0.85 & 0.87 & 0.88 \\
			\hline
		\end{tabular}
	\end{center}
\end{table}

We consider the configurations related to power control and mobility. Specifically, fours kinds of misconfigurations are considered: transmission power too strong, transmission power too weak, handover margin too large, and handover margin too small. These misconfigurations lead to severe interference and coverage gap, which are the main causes for link failures. As a realistic communication environment, we consider a two-tier HetNets comprised of multiple femtocells overlaid on a macrocell. Support Vector Machines (SVM) is used as the basic learner for model construction. The indoor and outdoor propagation models are determined based on the ITU and COST231 models. We compare the performance of the three architectures under different cell densities in Table~\ref{tab:sim}. {We observe that the accuracies of the cell-specific architecture keep almost unchanged while the other architectures yield higher accuracies with higher cell densities, in that there are more historical records for model training. As the transfer learning based architecture weighs historical instances from other cells according to the similarity between the target cell and these cells, it can effectively extract common knowledge from other cells' instances while eliminating misleading ones. The unified model based architecture, on the other hand, treats all instances equally, and thus can be misled by dissimilar instances.}


\section{Concluding Remarks}
This article has envisioned the crucial role of heterogeneity awareness in configuring and diagnosing ultra-dense small cells. Instead of assuming homogeneous usage patterns, heterogeneity awareness exploits heterogeneous traffic patterns, mobility behaviors and geospatial properties. Through careful investigation of fine-grained large-scale traces of millions of subscribers, we highlight the significance of heterogeneity awareness in small cell configurations and diagnosis. We further discuss the pros and cons of possible models and propose a heterogeneity-aware architecture. Some observations in the empirical study and the discussion of technical solutions can provide some implications for future designs of ultra-dense small cell configurations and diagnosis.

\section*{Acknowledgment}
The research was supported in part by the National Science Foundation of China under Grant 61502114, 91738202, 61729101, 61531011, and 61631015, the RGC under Contract CERG 16212714, 16203215, ITS/143/16FP-A, Major Program of National Natural Science Foundation of Hubei in China with Grant 2016CFA009, 2015ZDTD012.

\begin{biography}{Wei Wang (S'10-M'16)} (gswwang@connect.ust.hk) is currently a professor in School of Electronic Information and Communications, Huazhong University of Science and Technology. He received his Ph.D. degree in Department of Computer Science and Engineering from Hong Kong University of Science and Technology. He served as guest editors of Wireless Communications and Mobile Computing, IEEE COMSOC MMTC Communications, and TPC of INFOCOM, GBLOBECOM, etc. His research interests include PHY/MAC designs, security, and mobile computing in cyber-physical systems.
\end{biography}

\begin{IEEEbiography}{Lin Yang}	is a Postdoctoral fellow in Hong Kong University of Science and Technology (HKUST). He received his Ph.D and M.Phil degree from HKUST. Before he joined HKUST, he obtained his B.S degree in Computer Science from South China University of Technology (SCUT) in June 2008. His research interest includes security design for mobile systems and big data analaysis in mobile networks.
\end{IEEEbiography}

\begin{biography}{Qian Zhang (M'00-SM'04-F'12)} (qianzh@cse.ust.hk) joined Hong Kong University of Science and Technology in Sept. 2005 where she is a full Professor in the Department of Computer Science and Engineering. Before that, she was in Microsoft Research Asia, Beijing, from July 1999, where she was the research manager of the Wireless and Networking Group. Dr. Zhang has published about 300 refereed papers in international leading journals and key conferences in the areas of wireless/Internet multimedia networking, wireless communications and networking, wireless sensor networks, and overlay networking. She is a Fellow of IEEE for ``contribution to the mobility and spectrum management of wireless networks and mobile communications". Dr. Zhang has received MIT TR100 (MIT Technology Review) worlds top young innovator award. She also received the Best Asia Pacific (AP) Young Researcher Award elected by IEEE Communication Society in year 2004. Her current research is on cognitive and cooperative networks, dynamic spectrum access and management, as well as wireless sensor networks. Dr. Zhang received the B.S., M.S., and Ph.D. degrees from Wuhan University, China, in 1994, 1996, and 1999, respectively, all in computer science.
\end{biography}

\begin{biography}{Tao Jiang (M'06-SM'10)} (tao.jiang@ieee.org) is currently a Chair Professor with the School of Electronics Information and Communications, Huazhong University of Science and Technology, Wuhan, P. R. China. He has authored or co-authored more 300 technical papers in major journals and conferences and 5 books in the areas of wireless communications and networks. He was invited to serve as TPC Symposium Chair for the IEEE GLOBECOM 2013, IEEEE WCNC 2013 and ICCC 2013. He is served or serving as associate editor of some technical journals in communications, including in IEEE Transactions on Signal Processing, IEEE Communications Surveys and Tutorials, IEEE Transactions on Vehicular Technology, IEEE Internet of Things Journal, and he is the associate editor-in-chief of China Communications, etc.. He is a senior member of IEEE.
\end{biography}
\end{document}